\documentclass[12pt]{iopart}
\usepackage{graphicx}
\usepackage{epstopdf}
\usepackage{amssymb}
\usepackage{dcolumn}
\usepackage{iopams}
\usepackage{citesort}
\usepackage{datetime}

\begin{document}

\title[Magnetic and superconducting phase diagram of HoPdBi]{Magnetic and superconducting phase diagram of the half-Heusler topological semimetal HoPdBi}

\author{A M Nikitin$^1$, Y Pan$^1$, X Mao$^{1}$\footnote{Permanent address: Jimei University, Yinjiang Rd 185, 361021 Jimei, Xiamen, China}, R Jehee$^1$, G K Araizi$^1$, Y K Huang$^1$, C Paulsen$^2$, S C Wu$^3$, B H Yan$^3$ and A de Visser$^1$}

\address{$^1$ Van der Waals - Zeeman Institute, University of Amsterdam, Science Park 904, 1098 XH Amsterdam, The Netherlands}
\address{$^2$ Institut N\'{e}el, CNRS, and Universit\'{e} Joseph Fourier - BP 166, 38042 Grenoble, France}
\address{$^3$ Max-Planck-Institut f\"{u}r Chemische Physik fester Stoffe - N\"{o}thnitzer Strasse 40, 01187 Dresden, Germany}
\eads{\mailto{a.nikitin@uva.nl}, \mailto{a.devisser@uva.nl}}

\begin{abstract} We report a study of the magnetic and electronic properties of the non-centrosymmetric half-Heusler antiferromagnet HoPdBi ($T_N = 2.0$~K). Magnetotransport measurements show HoPdBi has a semimetallic behaviour with a carrier concentration $n=3.7 \times 10^{18}$~cm$^{-3}$ extracted from the Shubnikov-de Haas effect. The magnetic phase diagram  in the field-temperature plane has been determined by transport, magnetization and thermal expansion measurements: magnetic order is suppressed at $B_M\sim$ 3.6 T for $T \rightarrow 0$. Superconductivity with $T_c \sim 1.9$~K is found in the antiferromagnetic phase. Ac-susceptibility measurements provide solid evidence for bulk superconductivity below $T_c = 0.75$~K with a screening signal close to a volume fraction of  100 \%. The upper critical field shows an unusual linear temperature variation with $B_{c2}(T \rightarrow 0) = 1.1$~T. We also report electronic structure calculations that classify HoPdBi as a new topological semimetal, with a non-trivial band inversion of 0.25~eV.

\vspace{0.8cm}

\end{abstract}

\date{today}

\maketitle

\section{Introduction}

The compound HoPdBi is member of the large family of half-Heuslers that crystallize in the non-centrosymmetric space group $F\overline{4}3m$~\cite{Haase2002}. Ternary Heuslers with composition 2:1:1 and half Heuslers with composition 1:1:1 attract ample attention because of their flexible electronic structure, which gives rise to applications as multifunctional materials in, for example, the fields of spintronics and thermoelectricity~\cite{Graf2011}. Recently, a new incentive to investigate half-Heusler compounds was provided by first principles calculations~\cite{Lin2010,Chadov2010,Feng2010}. By employing the hybridization strength and the magnitude of the spin-orbit coupling is was established that "heavy-element" half-Heusler compounds may show a non-trivial band inversion, similar to the prototypical topological insulator HgTe~\cite{Bernevig2006,Hasan&Kane2010,Qi&Zhang2010}. Topological half-Heuslers are zero-gap semiconductors, where the topological surface states can be created by applying strain that opens the gap. The emergence of magnetic order and superconductivity at low temperatures in half-Heuslers provides a new research direction to study the interplay of topological states, superconductivity and magnetic order~\cite{Pan2013}.

Recently, much research has been devoted to the band-inverted half-Heusler platinum and palladium bismuthides, notably because some of these superconduct at low temperatures~\cite{Yan&deVisser2014}: LaPtBi ($T_c = 0.9$~K~\cite{Goll2008}), YPtBi ($T_c = 0.77$~K~\cite{Butch2011,Bay2012b}), LuPtBi ($T_c = 1.0$~K~\cite{Tafti2013}) and LuPdBi ($T_c = 1.7$~K~\cite{Xu2014,Pavlosiuk2015}). The $s$-$p$ inverted band order ($\Gamma _8 > \Gamma _6$) makes these compounds candidates for topological superconductivity~\cite{Kitaev2009,Schnyder2009}. Topological superconductors are predicted to be unconventional superconductors, with an admixture of even- and odd-parity Cooper pair states in the bulk, and protected Majorana states at the surface~\cite{Hasan&Kane2010,Qi&Zhang2010}. Since the crystal structure lacks inversion symmetry these superconductors might also be classified as non-centrosymmetric. Here theory predicts a mixed even- and odd-parity Cooper pair state~\cite{Frigeri2004} as well.

Recently, we reported the observation of superconductivity ($T_c = 1.22$~K) and antiferromagnetic order ($T_N = 1.06$~K) in the half-Heusler ErPdBi. In addition, electronic structure calculations predicted a sizeable band inversion, which makes ErPdBi a first candidate for the investigation of the interplay of topological states, superconductivity and magnetic order. Another candidate proposed is CePdBi ($T_c = 1.3$~K, $T_N = 2.0$~K), however, here superconductivity develops in a small sample volume and is associated with a disordered phase~\cite{Goraus2013}. The coexistence of superconductivity and magnetic order in ErPdBi provided a strong motivation to search for similar phenomena in other REPdBi compounds (RE = Rare Earth), such as HoPdBi~\cite{Haase2002,Gofryk2005}. HoPdBi was reported to be an antiferromagnet with a N\'{e}el temperature $T_N = 2.2$~K~\cite{Gofryk2005}. Its magnetic susceptibility follows a Curie-Weiss law with an effective moment $\mu_{\rm {eff}}$ = 10.6 $\mu_B$, in good agreement with the theoretical value for free RE$^{3+}$ ions with Russell-Saunders coupling ($J=8$). The weak antiferromagnetic coupling is furthermore inferred from the low value of the Curie-Weiss constant $\Theta _P = - 6.4$~K. Here we report results of transport, magnetic and thermal measurements on single crystalline HoPdBi. Our flux-grown crystals show antiferromagnetic order at $T_N = 2.0$~K and superconductivity at $ T_c \sim 1.9$~K. However, the transition to bulk superconductivity, inferred from the full diamagnetic screening signal (100 \% volume fraction), sets in near $ T_c ^{bulk} \sim 0.75$~K. We report the magnetic and superconducting phase diagram in the field-temperature plane, as well as electronic structure calculations that reveal a non-trivial band inversion of 0.25~eV at the $\Gamma$-point. In the course of our investigations, Nakajima \textit{et al}.~\cite{Nakajima2015} reported that superconductivity occurs in almost the entire antiferromagnetic REPdBi series, with $T_c \sim 1$~K for HoPdBi.

\section{Methods}

Several single crystalline batches of HoPdBi were prepared using Bi flux. The high purity elements Ho, Pd and Bi (purity 3N5, 4N and 5N, respectively) served as starting material. First an ingot of HoPdBi was prepared by arc-melting and placed in an alumina crucible with excess Bi flux. The crucible was sealed in a quartz tube under a high-purity argon atmosphere (0.3 bar). Next the tube with crucible was heated in a furnace to 1150 $^{\circ}$C and kept at this temperature for 36 h. Then the tube was slowly cooled to 500 $^{\circ}$C at a rate of 3 $^{\circ}$C per hour to form the crystals. The single-crystalline nature of the crystals was checked by Laue backscattering. After cutting the crystals in suitable shapes for the various experiments, the surfaces were cleaned by polishing. Powder X-ray diffraction confirmed the F$\bar{4}$3m space group and the extracted lattice parameter, $a$ = {6.605} {\AA}, is in good agreement with the literature~\cite{Haase2002}. Several tiny extra diffraction lines point to the presence of a small amount of an impurity phase ($< 5$~\%). These lines could not be matched to potential inclusions or impurity phases, such as Bi, PdBi, $\alpha$-Bi$_2$Pd and $\beta$-Bi$_2$Pd. The results presented here are taken on samples that all come from the same single-crystalline batch. The measured samples are labeled \#2-\#4, and show very similar magnetic and superconducting properties.

Magnetic and transport measurements were carried out in a Physical Property Measurements System (PPMS, Quantum Design) in the temperature range 1.8-300~K. The magnetization and susceptibility were measured utilizing a Vibrating Sample Magnetometer, while the resistance and Hall effect were measured using a standard four-point low-frequency ac-technique with an excitation current of 1 mA. The specific heat was measured down to 2.0 K on a crystal with a mass of 10 mg in the PPMS as well. Low temperature, $T = 0.24 - 10$~K, resistance and ac-susceptibility measurements were made in a $^3$He refrigerator (Heliox, Oxford Instruments) using a low-frequency lock-in technique with low excitation currents ($I \leq 200$~$\mu$A). The coefficient of linear thermal expansion, $\alpha$ = L$^{-1}$(dL/dT), with $L$ the sample length, was measured using a three-terminal parallel-plate capacitance method using a home-built sensitive dilatometer based on the design reported in Ref.~\cite{Schmiedeshoff2006}. Magnetoresistance data were taken in a dilution refrigerator (Kelvinox, Oxford Instruments) for $T = 0.03 - 1$~K and high magnetic fields up to 17 T.  Additional low-temperature dc-magnetization and ac-susceptibility measurements were made using a SQUID magnetometer, equipped with a miniature dilution refrigerator, developed at the N\'{e}el Institute.

\section{Results and analysis}

\subsection{Superconductivity}

\begin{figure}
\begin{center}
\includegraphics[height=6cm]{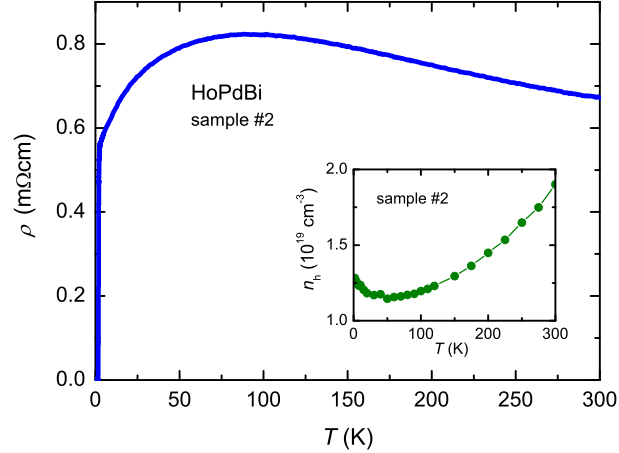}
\caption{Resistivity and carrier concentration (inset) as a function of temperature of HoPdBi sample~\#2.}
\end{center}
\end{figure}

\begin{figure}
\begin{center}
\includegraphics[height=6cm]{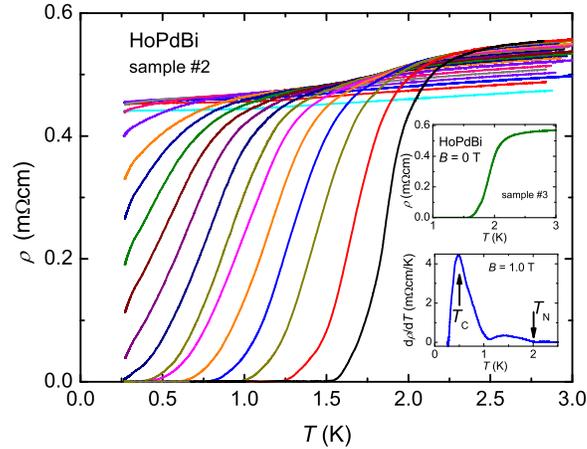}
\caption{Resistivity as a function of temperature for HoPdBi sample~\#2 measured in fixed magnetic fields: from right to left 0~T to 1.1~T in steps of 0.1~T, 1.2~T to 2~T in steps of 0.2~T, and 2.25~T, 2.5~T and 3~T. Lower inset: Derivative d$\rho$/d$T$ \textit{versus} $T$ for $B = 1.0$~T; arrows indicate $T_c$ and $T_N$. Upper inset: $\rho (T)$ \textit{versus} $T$ in zero field for sample~\#3.}
\end{center}
\end{figure}

In fig.~1 we show the resistivity $\rho(T)$ of a single crystal of HoPdBi (sample \#2). The resistivity values are rather large and $\rho(T)$ has a broad maximum around 80 K. The charge carrier concentration $n_h$ extracted from the low field ($B < 2$~T) Hall effect measurements is traced in the inset of fig.~1. The carriers are holes and at the lowest temperature ($T=2.0$~K) $n_h = 1.3\times 10^{19}$~cm$^{-3}$. These transport parameters reveal semimetallic-like behaviour. At low temperatures, the resistivity drops to zero, which signals the superconducting transition.

The superconducting transition measured by resistivity in zero and applied magnetic fields is reported in detail in fig.~2. The superconducting transition at $B=0$ is fairly broad, with an onset temperature of 2.3~K and zero resistance at 1.6~K. By tracing the maximum in the derivative d$\rho$/d$T$ we obtain $T_c$ = 1.86 K. The broad transition is most likely related to the simultaneous development of antiferromagnetic order with $T_N = 2.0$~K (see below). A similar rounded $\rho(T)$ near the onset for superconductivity was observed for ErPdBi, where $T_c \simeq T_N$ as well~\cite{Pan2013}. We have determined the upper critical field $B_{c2}$ by taking $T_c (B)$ as the maximum in d$\rho$/d$T$ obtained in constant magnetic fields. See the inset of fig.~2 for the data at $B = 1.0$~T. Here the structure at 2.0 K signals the antiferromagnetic transition. The temperature variation $B_{c2}(T)$ is reported in the phase diagram, fig.~7. In the upper inset of fig.~2 we show the zero-field data for sample \#3. The resistivity data for this sample (data in field are not shown)  are in good agreement with the results obtained for sample \#2.

\begin{figure}
\begin{center}
\includegraphics[height=6cm]{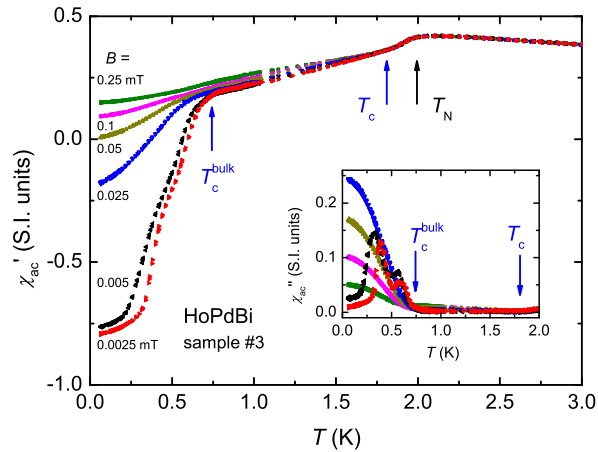}
\caption{Real part of the ac-susceptibility of HoPdBi sample \#3 for driving fields of 0.0025, 0.005, 0.025, 0.05, 0.10 and 0.25~mT, from bottom to top. Arrows indicate $T_N$, $T_c$ and $T_c ^{bulk}$. The inset shows the corresponding imaginary part of $\chi_{ac}$. }
\end{center}
\end{figure}

In fig.~3 we show ac-susceptibility ($\chi _{ac}$) data taken on sample \#3 for different driving fields $B_{ac}$ ranging from $0.0025$ to $0.25$~mT. Upon cooling $\chi _{ac}(T)$ first increases and has a pronounced maximum at 2.0~K, which locates the antiferromagnetic phase transition temperature. Upon further cooling a large superconducting signal appears below 0.75~K. For the smallest driving fields $B_{ac} \leq 0.05$~mT the screening fraction associated with this diamagnetic signal is $\sim$~90~\%, which provides strong support for bulk superconductivity.  Upon increasing the driving field to 0.25~mT the magnitude of the diamagnetic signal is rapidly depressed, which indicates the lower critical field $B_{c1}$ is very small. However, a close inspection of the data shows that a weaker diamagnetic signal develops near 1.8 K, which is close to midpoint of the superconducting transition at 1.9 K of sample \#3 (see upper inset fig.~2). The amplitude of the diamagnetic signal varies with $B_{ac}$. This clearly indicates a smaller volume fraction ($\sim 20$~\%) of the sample becomes superconducting at 1.8 K. Here we have neglected the demagnetization factor that we estimate to be $N = 0.07 \pm 0.02$ for our needle-shaped sample \#3. We remark that after correcting for demagnetization effects the total superconducting volume fraction is close to $\sim 100$~\%. These observations are corroborated by the imaginary part of $\chi_{ac}(T)$ shown in the inset of fig.~3. The transition to the bulk superconducting state reveals the presence of a remaining sample inhomogeneity, as indicated by the double peak structure in $\chi _{ac} ^{\prime \prime}$ for small values $B_{ac} \leq 0.005$~mT.

\subsection{Antiferromagnetic order}

Magnetic susceptibility measurements carried out in a field of 0.1~T confirm local moment behaviour of the Ho moments. For $T > 50$~K Curie-Weiss behaviour is observed with an effective moment $\mu_{eff} = 10.3~\mu_B$ and a negative Curie-Weiss constant $\Theta_P = -5.7$~K indicating the presence of antiferromagnetic interactions. These values are close to the ones reported in Ref.~\cite{Gofryk2005}. In fig.~4 we show the dc-susceptibility around the antiferromagnetic transition measured in fields up to 3~T down to very low temperatures (0.1 K). The N\'{e}el temperature is progressively shifted with field and can no longer be discerned at the largest field. The variation $T_N (B)$ is traced in the phase diagram (fig.~7). In the right panel we show the magnetization measured at $T= 0.075$~K, deep in the antiferromagnetic phase, and at 2~K and 5~K, measured along the [100] axis. At the lowest temperature the magnetization increases rapidly until it reaches a value of $\sim 6~ \mu _B$ at 3~T after which the increase is more gradual. In a field of 8~T the magnetization attains a value of $7.2~\mu _B$ which is still considerably below the saturation value $m_s = gJ = 10 ~\mu_B$ (here $g = 5/4$ is the Land\'{e} factor). An explanation for this is offered by the sizeable magnetic anisotropy associated with the Ho ions (see section 4). The presence of magnetocrystalline anisotropy can also be inferred from the very weak signature of a metamagnetic-like transition observed in $M(B)$ below $T_N$.

\begin{figure}
\begin{center}
\includegraphics[height=6cm]{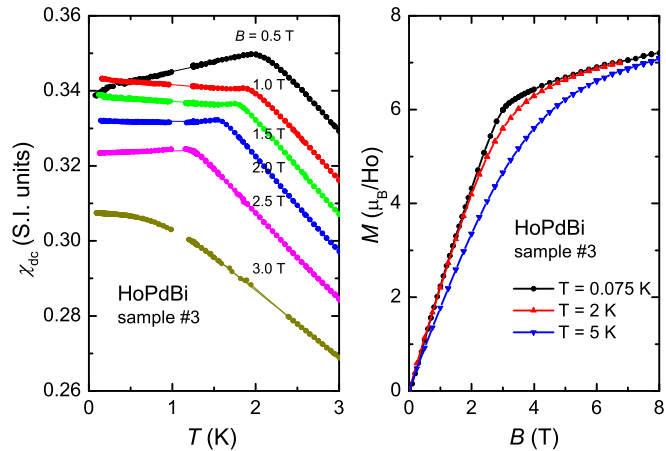}
\caption{Left panel: Magnetic susceptibility of HoPdBi sample \#3 as a function of temperature measured in magnetic fields up to 3~T as indicated. Right panel: Magnetization in fields up to 8~T at temperatures of 0.075, 2.0 and 5.0 K.}
\end{center}
\end{figure}

\begin{figure}
\begin{center}
\includegraphics[height=6cm]{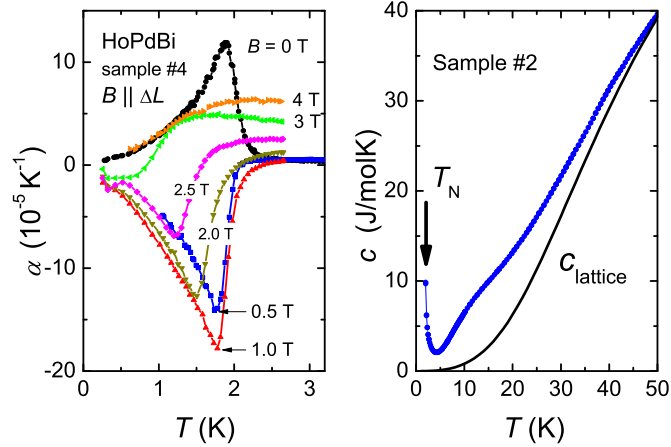}
\caption{Left panel: Coefficient of linear thermal expansion as a function of temperature of HoPdBi (sample \#4) measured in zero and applied fields as indicated. Right panel: Specific heat (sample \#2) in a plot of $c$ versus $T$ up to 50~K. The solid line indicates the lattice contribution given by the Debye function with $\theta _D = 193$~K.}
\end{center}
\end{figure}

In fig.~5 we show the coefficient of linear thermal expansion $\alpha (T)$ around the antiferromagnetic transition. Upon cooling a large positive step $\Delta \alpha$ is observed, where the midpoint nicely agrees with $T_N$ determined by the magnetic susceptibility. The superconducting transition is not observed in the thermal expansion because of the relatively small entropy involved (see the discussion section). In the right panel we show the specific heat measured in the temperature range 2-50 K. The increase at low temperatures upon approaching 2 K is due to the antiferromagnetic transition. When we compare $c(T)$ with the lattice specific heat, estimated by a Debye function with a Debye temperature $\theta _D = 193$~K, a broad magnetic contribution centered around 15~K becomes noticeable. This magnetic contribution with an entropy of $ \sim 1.7 \times R$, where $R=8.31$ J/molK is the gas constant, is most likely related to the partly lifting of the 17-fold degenerated crystalline electric field ground state in the cubic symmetry. We remark that the magnetic susceptibility starts to deviate from the Curie-Weiss behaviour below 50~K as well. By combining the thermal expansion and specific heat data we can make an estimate of the pressure variation of $T_N$ using the Ehrenfest relation d$T_N / $ d$p = V_m \Delta \beta / (\Delta c/T_N)$. Here the coefficient of volume expansion $\beta = 3 \alpha$ and $V_m = 4.3 \times 10^{-5}$~m$^3$/mol is the molar volume. Using the experimental values $\Delta \alpha = 1.2 \times 10^{-4}$~K$^{-1}$ and $\Delta c/T_N \sim$ 5~J/molK$^2$ we obtain d$T_N / $ d$p \sim 0.3$~K/kbar. In an applied field, the longitudinal thermal expansion measured along the direction of the magnetic field, $\alpha _\parallel$, develops a pronounced minimum. The negative $\alpha _\parallel$ indicates a large magnetocrystalline anisotropy, with an expansion along the field and a contraction perpendicular to the field, upon cooling below $T_N$. Notice, in magnetic field $ \beta = \alpha_\parallel + 2 \alpha _\perp$. The variation $T_N (B)$ extracted from the thermal expansion is traced in the phase diagram fig.~7.

\subsection{Quantum oscillations}

The magnetoresistance was measured in magnetic fields up to 17~T at low temperatures in the dilution refrigerator. A typical trace taken at $T= 0.05$~K is shown in fig.~6. After the suppression of superconductivity, the kink at $B_M = 3.6$~T locates the antiferromagnetic phase boundary. In fields exceeding 8~T we observed Shubnikov-de Haas oscillations. Since the oscillatory component is rather small ($\Delta \rho / \rho \sim 0.1 \%$) and superimposed on a non-monotonic background, it is difficult to extract the precise magnitude as a function of $1/B$. In the lower inset of fig.~6 we trace the resulting oscillations in $\Delta \rho$ as a function of $1/B$. A fast Fourier transform (upper inset) shows the frequency is $75 \pm 5$~T. For a circular extremal cross section $A_k (E_F) = \pi k_F^2$ we calculate with help of the Onsager relation a Fermi wave vector $k_F = 4.8\times 10^8$~m$^{-1}$. Assuming a spherical Fermi surface pocket, the corresponding 3D carrier density $n=k_F^3/3\pi^2 = 3.7\times 10^{18}$~m$^3$ which is a factor $\sim 3.5$ smaller than the value deduced from the Hall effect measurements. Additional measurements at $T= 0.3$ and 0.6~K show the thermal damping is very weak, which points to an effective mass much smaller than the free electron mass, $m_{eff} < 0.5 m_e$, as reported for other topological semimetals~\cite{Butch2011,Goll2002}.

\begin{figure}
\begin{center}
\includegraphics[height=6cm]{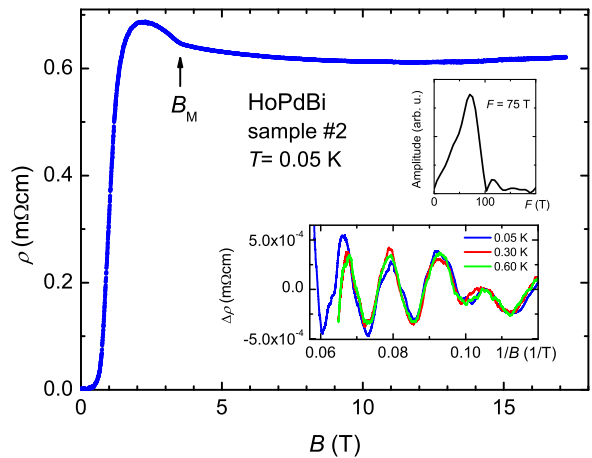}
\caption{Resistance of HoPdBi sample \#2 as a function of the magnetic field at $T=0.05$~K. The arrow indicates the antiferromagnetic phase boundary at $B_M$. Lower inset: $\Delta \rho$ as a function of $1/B$ showing the Shubnikov - de Haas effect at 0.05, 0.3 and 0.6 K. Upper inset: fast Fourier transform of the data with frequency $F= 75 \pm 5$~T. }
\end{center}
\end{figure}

\subsection{Phase diagram}

In fig.~7 we present the magnetic and superconducting phase diagram of HoPdBi. The upper critical field $B_{c2}$ shows a smooth variation with temperature and extrapolates to the value of 1.1~T at $T=0$. This tells us $B_{c2}(T)$ is associated with the superconducting transition observed at 1.9~K in zero field. When we compare the $B_{c2}$-data to the Werthamer-Helfand-Hohenberg (WHH) curve for a weak-coupling spin-singlet orbital-limited superconductor, some departure appears below $\sim 0.7$~K. We remark this is close to the temperature below which the diamagnetic screening signal shows the sudden increase towards bulk superconductivity (see fig.~2). In the WHH model the clean-limit zero-temperature orbital critical field can be calculated from the relation $B_{c2}^{orb}(0) = 0.72 \times T_c |$d$B_{c2}/$d$T|_{T_c}$ and attains a value of 0.9 T (in the dirty limit the coefficient 0.72 is replaced by 0.69). From the experimental value $B_{c2}(0)=1.1$~T we calculate, utilizing the relation $B_{c2} = \Phi _0 / 2 \pi \xi ^2$, where $\Phi _0$ is the flux quantum, a superconducting coherence length $\xi = 17$~nm. A departure from the WHH model has also been observed for ErPdBi~\cite{Pan2013} and notably for YPtBi~\cite{Butch2011,Bay2012b}, where is was taken as a strong indication of an odd-parity component in the superconducting order parameter. We remark, the superconducting phase is completely embedded in the antiferromagnetic phase.

The antiferromagnetic phase boundary has been determined by resistivity, magnetization and thermal expansion measurements, see fig.~7. After an initial ($B < 1$~T) steep increase with field, $T_N$ smoothly decreases, and the antiferromagnetic phase transition is suppressed at $B_M (0)= 3.6$~T. Values for $T_N$ extracted from the magnetization and thermal expansion are somewhat lower than the values obtained by transport, but all three experiments track the same phase boundary. These differences might be partly attributed to different orientations of the crystals in the magnetic field in connection to the large magnetic anisotropy. For the magnetization experiments the magnetic field was applied along [100],  while for the other experiments the field was applied along an arbitrary crystal direction. In the figure we compare the phase boundary with the phenomenological order parameter function $B_M (T) = B_M (0)(1-(T/T_N)^{\alpha})^{\beta}$ with $T_N = 2.0$~K, $B_M (0) =3.6$~T, $\alpha = 2$ and $\beta = 0.34$. The latter value is close to the theoretical value $\beta = 0.38$ expected for a 3D Heisenberg antiferromagnet~\cite{Domb1996}.

\begin{figure}
\begin{center}
\includegraphics[height=6cm]{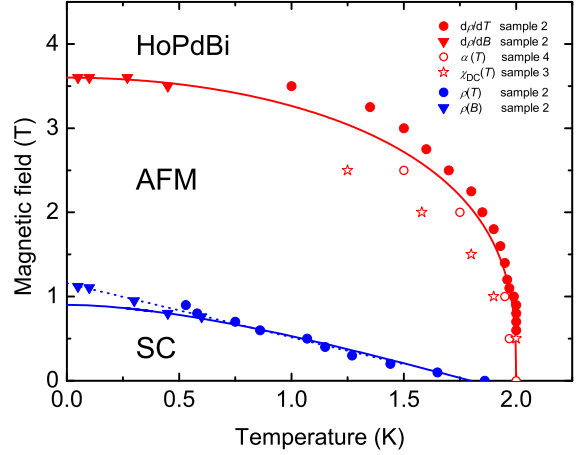}
\caption{Magnetic and superconducting phase diagram of HoPdBi.
Data points for the superconducting (SC) transition are taken on sample \#2 where $T_c (B)$ is defined by the temperature of the maximum in d$\rho /$d$T$ at fixed fields (blue circles) or the maximum in d$\rho /$d$B$ at fixed temperatures (blue triangles).
The blue solid line is a comparison of the $B_{c2}(T)$ data with the WHH model (see text). The blue dashed line represents a linear fit to the data. Data points for the antiferromagnetic (AFM) transition are taken from the resistivity (sample \#2: red circles and triangles from d$\rho /$d$T$ and d$\rho /$d$B$, respectively), the magnetization (sample \#3: open stars) and thermal expansion (sample \#4: open circles). The red solid line represents a phenomenological order parameter fit with $T_N = 2.0$~K and $B_M = 3.6$~T at $T=0$ (see text)}.
\end{center}
\end{figure}

\subsection{Electronic structure}

\begin{figure}
\begin{center}
\includegraphics[height=7cm]{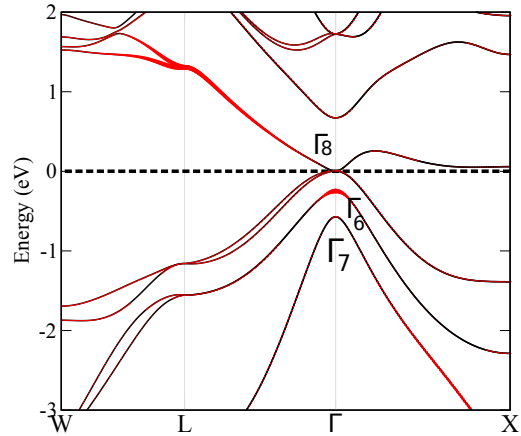}
\caption{Calculated bulk band structure of the half-Heusler HoPdBi. The projection of the Pd-$s$ states is highlighted by the size of the filled red circles. Bands $\Gamma _{6,7,8}$ are labeled according to the symmetry of their wave functions. The Fermi energy, $E_F$, is shifted to zero (horizontal dashed line). The measured sample is slightly $p$-doped, which will cause $E_F$ to lie marginally below zero.}
\end{center}
\end{figure}

To understand the electronic properties, we have performed \textit{ab initio} band structure calculations~\cite{Kresse&Furthmuller1996} based on the density-functional theory. The core electrons are represented by the projector-augmented-wave potential. We keep only three valence electrons of Ho (equivalent to La) and freeze all other $f$ electrons as core electrons, to simplify the understanding of the topology. The calculated bulk band structure of HoPdBi exhibits a zero-gap, and is quite similar to that of ErPdBi~\cite{Pan2013} and other half-Heusler topological semimetals ~\cite{Lin2010,Chadov2010}. The $\Gamma_6$ band (contributed by Pd-$s$ states) is below the $\Gamma_8$ bands, showing the hall mark of band inversion. Therefore, we can clearly conclude that HoPdBi is also a topological semimetal. The effect of magnetization of Ho-$f$ states is expected to possibly split the spin degeneracy slightly, but to preserve the inverted band order because the band inversion strength (the energy difference between $\Gamma_8$ and $\Gamma_6$ bands) is as large as $0.25$~eV. The measured sample is slightly $p$-doped, which will cause $E_F$ to lie marginally below zero.

\section{Discussion}
Our systematic transport, magnetic and thermal properties study demonstrates that HoPdBi combines local moment magnetism and superconductivity and thus behaves very similar to the related half-Heusler compound ErPdBi~\cite{Pan2013}. Superconductivity occurs at $T_c = 1.9$~K as determined by the resistive transition and the onset in $\chi _{ac}$, however, the transition to a full diamagnetic screening fraction sets in at a lower temperature, near 0.75~K. These results are in line with the recent work by Nakajima \textit{et al}.~\cite{Nakajima2015}, but nevertheless there are some noticeable differences: the midpoint of the resistive transition is at $\sim 1.0$~K, and the onset of the diamagnetic signal is at $\sim 0.9~$~K ~\cite{Nakajima2015}. Moreover, the transition measured by $\chi _{ac}$ ~\cite{Nakajima2015} is very broad and extends to the lowest temperature measured ($\sim 0.1$~K). A comparison of both studies suggests the superconducting phase transition found below $\sim 0.9~$~K by Nakajima \textit{et al}.~\cite{Nakajima2015} is to be associated with the transition to the bulk superconducting state at 0.75 K as reported in fig.~2. The presence of a small volume fraction of our sample that superconducts at 1.9~K remains a puzzling aspect. A possible explanation is the presence of an impurity phase. Among the binary Bi-Pd alloys $\alpha$-Bi$_2$Pd is reported to superconduct at 1.7~K~\cite{Matthias1963}. However, the few extra tiny lines in the diffraction pattern could not be matched to the structure of $\alpha$-Bi$_2$Pd. Moreover, $B_{c2}(0)$ of $\alpha$-Bi$_2$Pd is small~\cite{Nakajima2015}, which is at variance with the measured value $B_{c2}(0) = 1.1$~T.

A second, appealing explanation for superconductivity at 1.9~K is the occurrence of surface superconductivity associated with the topological surface states. Surface superconductivity at temperatures higher than the bulk $T_c$ was recently proposed by Kapitulnik \textit{et al.}~\cite{Kapitulnik2014,Kapitulnik2015} for the topological half-Heuslers YPtBi and LuPtBi. Experimental evidence for this was provided by (surface sensitive) Scanning Tunneling Spectroscopy measurements. Surface superconductivity may also offer an explanation for the "unknown impurity phase" with $T_c = 1.7$~K, $i.e.$ above the bulk $T_c$ of 1.2~K, reported for ErPdBi~\cite{Pan2013}. Interestingly, Nakajima \textit{et al}.~\cite{Nakajima2015} claim that surface superconductivity with $T_c = 1.6$~K can be induced by heat treatments at 200~$^{\circ}$C for all REPdBi. These results may open a new, fascinating research direction in the field of topological superconductivity. Future experiments have to be directed to disentangle surface and bulk superconductivity in order to put these ideas on a firmer footing.

Local-moment antiferromagnetic order has been detected for most members of the REPdBi series~\cite{Riedemann1996,Gofryk2005,Pan2013,Nakajima2015}, with $T_N$ following the De Gennes scaling: $T_N \propto (g_J -1)^2 J(J+1)$ with $g_J$ the Land\'{e} factor~(see \textit{e.g.} \cite{Jensen&Mackintosh1991}). The Ne\'{e}l temperature $T_N = 2.0$~K of our crystals is in between the values 2.2~K and 1.9 K reported in Refs.~ \cite{Gofryk2005} and~\cite{Nakajima2015}, respectively. The phase diagram (fig.~7) shows that antiferromagnetic order entirely encloses the superconducting phase. The low-temperature $\chi _{ac}$ and dc-magnetization measurements indicate that  superconductivity and antiferromagnetic order coexist. In order to provide solid proof for this muon spin rotation/relaxation ($\mu$SR) and/or nuclear magnetic resonance experiments (NMR) would be very helpful. Alternatively, both ordering phenomena could possibly occupy different sample volumes. From recent neutron diffraction experiments~\cite{Nakajima2015} it was concluded that HoPdBi has a type-II antiferromagnetic structure: ferromagnetic planes couple antiferromagnetically with a propagation vector along the cube diagonal, $Q = (0.5, 0.5, 0.5)$. The  magnetization curves (fig.~4) and the thermal expansion in field (fig.~5) reveal the magnetocrystalline anisotropy is rather strong. This offers the following scenario for the magnetization curves. When the field is increased along the [100] axis the spins progressively reorient towards the field, while still pointing along the [111] direction. Near the antiferromagnetic phase boundary the component of the spin along the [100] direction is $m_s / \sqrt 3$ = 5.8~ $\mu_B$ which is close to the experimental value. For fields above 3~T the spins are being pulled against the anisotropy field, which explains the reduced value with respect to $m_s = 10 ~\mu _B$. Interestingly, theoretical work based on symmetry classifications predicts the type II antiferromagnetic structure in these half-Heuslers produces a new ground state: an antiferromagnetic topological insulator~\cite{Mong2010}. Recently, it was proposed that a first realization of this novel electronic state is found in the half-Heusler antiferromagnet GdPtBi ($T_N=9$~K)~\cite{Muller2014}.

The large contribution of antiferromagnetic order to the thermal properties (fig.~5) hampers the detection of superconductivity in the thermal expansion and specific heat. An estimate for the step-size $\Delta c$ in the specific heat at $T_c$ can be calculated from the standard BCS relation $(\Delta c /T_c )/\gamma = 1.43$, where $\gamma$ is the Sommerfeld coefficient. It appears that $\gamma$ cannot reliably be deduced from the experimental data (see fig.~5) because of the presence of the large magnetic contribution centered around 15~K. However, an estimate for $\gamma$ can be obtained from the Shubnikov - de Haas data using the relation $\gamma = \pi^2 k_B^2 g(E_F)/3$, where $g(E_F)=m^* k_F / \hbar^2 \pi^2$ is the density of states. With the parameters calculated in section 3.3 we obtain $\gamma < 0.1$~mJ/molK$^2$ and $\Delta c < 0.1$~mJ/K at $T_c$, which is very small indeed. An estimate for the step size in the thermal expansion coefficient, $\Delta \alpha$, at $T_c$ can be obtained from the Ehrenfest relation d$T_c /$d$p = V_m 3 \alpha / (\Delta c/T_c)$. Assuming a typical value for d$T_c / $ d$p$ = 0.044 K/GPa~\cite{Bay2012b} we calculate $\Delta \alpha$ at $T_c$  is smaller than $4.5 \times 10^{-11}$~K$^{-1}$, which is beyond the resolution of the dilatometer.

The occurrence of superconductivity in a Ho-based compound is uncommon. Rare examples are found among the Chevrel phases~\cite{Shelton1983} and the heavy-rare earth cuprates~\cite{Hor1987} and pnictides~\cite{Yang2009}. Coexistence of superconductivity and antiferromagnetic order is found in the borocarbide HoNi$_2$B$_2$C~\cite{Cava1994}, but here $T_c = 10.5$~K and exceeds $T_N = 6.0$~K. An important new aspect in the REPdBi series is the non-centrosymmetric crystal structure and the topological nature of the electronic structure. This provides a new opportunity to investigate unconventional superconductivity, due to mixed even and odd parity Cooper pair states, and its interplay with long-range magnetic order.

\section{Summary}

We have investigated the transport, magnetic and thermal properties of the half-Heusler antiferromagnet HoPdBi. Electrical resistivity and ac-susceptibility data taken on flux-grown single crystals show superconductivity occurs below 1.9~K. We demonstrate HoPdBi is a bulk superconductor. However, the transition to bulk superconductivity, as indicated by a diamagnetic screening fraction close to 100 \%, sets in at a lower temperature of 0.75 K. The N\'{e}el temperature $T_N$ is $2.0$~K as determined by thermal expansion and dc-magnetization measurements. The superconducting and magnetic phase diagram in the $B-T$ plane has been determined: superconductivity is confined to the antiferromagnetic phase. Electronic structure calculations show HoPdBi is a topological semimetal with a band inversion of 0.25~eV at the $\Gamma$-point. This is in-line with the semimetallic behaviour observed in the electrical resistivity and the low carrier concentration $n_h = 3.7 \times 10^{18}$~cm$^{-3}$ extracted from the Shubnikov-de Haas effect. We conclude, HoPdBi belongs to the half-Heusler REPdBi series with a topological band structure and presents a new laboratory tool to study the interplay of antiferromagnetic order, superconductivity and topological quantum states~\cite{Pan2013,Nakajima2015}.

\ack{This work is part of the research programme on Topological Insulators of the Foundation for Fundamental Research on Matter (FOM), which is part of the Netherlands Organisation for Scientific Research (NWO). B.Y. acknowledges financial support from a European Research Council (ERC) Advanced Grant (291472).}

\section*{References}
\bibliography{RefsTI_all}

\end{document}